\documentclass[aps,superscriptaddress,prc,twocolumn,nofootinbib]{revtex4}

\usepackage{graphicx}
\usepackage{epstopdf}
\usepackage{subfigure}
\usepackage{epsfig}
\usepackage{amsmath,amssymb,amsfonts}
\usepackage{color}
\usepackage[utf8]{inputenc}
\allowdisplaybreaks
\usepackage[bookmarks,
                bookmarksopen = true,
                bookmarksnumbered = true,
                linktocpage,
                colorlinks = true,
                linkcolor = blue,
                urlcolor  = blue,
                citecolor = blue,
                anchorcolor = green,
                hyperindex = true,
                hyperfigures]
                {hyperref}
\usepackage{multirow}
\usepackage{array}

\begin{document}

\title{Reexamining charm versus bottom quark energy loss inside a color-deconfined medium}

\author{Yichao Dang}
\affiliation{Institute of Frontier and Interdisciplinary Science, Shandong University, Qingdao, Shandong 266237, China}

\author{Wen-Jing Xing}
\email{wenjing.xing@mails.ccnu.edu.cn}
\affiliation{Institute of Frontier and Interdisciplinary Science, Shandong University, Qingdao, Shandong 266237, China}

\author{Shanshan Cao}
\email{shanshan.cao@sdu.edu.cn}
\affiliation{Institute of Frontier and Interdisciplinary Science, Shandong University, Qingdao, Shandong 266237, China}

\author{Guang-You Qin}
\email{guangyou.qin@mail.ccnu.edu.cn}
\affiliation{Institute of Particle Physics and Key Laboratory of Quark and Lepton Physics (MOE), Central China Normal University, Wuhan, 430079, China}

\date{\today}


\begin{abstract}
The general intuition that heavier partons suffer weaker energy loss inside a quark-gluon plasma (QGP) medium is critically re-examined. Within a linear Boltzmann transport model that includes both Yukawa and string types of interactions between heavy quarks and the QGP, we find that while the radiative energy loss is suppressed by the parton mass, heavier partons can experience stronger string potential scatterings with the medium. Their competition may result in less energy loss of bottom quarks than charm quarks at low transverse momentum ($p_\mathrm{T}$) but an inverse order at high $p_\mathrm{T}$. Our model calculation shows a weaker nuclear modification on bottom particles than charm particles at low $p_\mathrm{T}$, as observed by both RHIC and LHC experiments, but predicts an opposite hierarchy at high $p_\mathrm{T}$. A larger momentum space transport coefficient ($\hat{q}$) and a smaller spatial diffusion coefficient ($D_\mathrm{s}$) are found for bottom quarks than for charm quarks.
\end{abstract}

\maketitle


\section{Introduction}
\label{sec:Introduction}

High-energy nuclear collisions conducted at the Relativistic Heavy-Ion Collider (RHIC) and the Large Hadron Collider (LHC) provide a unique opportunity for studying Quantum Chromodynamics (QCD) at extremely high temperature and high density. It is now generally accepted that a strongly coupled color de-confined matter, known as quark-gluon plasma (QGP), is created in these energetic collisions, which behaves like a perfect fluid~\cite{Gyulassy:2004zy,Jacobs:2004qv}. Among various probes of the QGP, heavy quarks (charm and bottom quarks) are of particular interest. Due to their large masses, they are mainly produced from the primordial hard scatterings between nucleons and then scatter through the QGP with their flavors conserved. Therefore, one may infer properties of the QGP by comparing the spectra of heavy flavor particles between nucleus-nucleus collisions and proton-proton collisions~\cite{Dong:2019byy,Dong:2019unq}. 

Heavy flavor phenomenologies at different transverse momentum ($p_\mathrm{T}$) scales are driven by different interaction mechanisms. At high $p_\mathrm{T}$, heavy quarks lose energy inside the QGP mainly through inelastic scattering, or medium-induced gluon emission process~\cite{Zhang:2003wk,Abir:2015hta,Zhang:2018nie,Du:2018yuf}. This is similar to the evolution of energetic light flavor partons, and therefore can be described using the same framework developed for jet quenching with mass effects properly taken into account~\cite{Buzzatti:2011vt,Djordjevic:2013pba,Xu:2014ica,Xu:2015bbz,Cao:2017hhk,Xing:2019xae,Ke:2020clc}. At low $p_\mathrm{T}$, the phase space for the gluon bremsstrahlung can be significantly suppressed by the large mass of heavy quarks, known as the ``dead cone" effect~\cite{Dokshitzer:2001zm,Abir:2012pu,ALICE:2021aqk}, and the heavy quark motion is then dominated by their quasi-elastic scatterings with the QGP~\cite{Moore:2004tg,Mustafa:2004dr,Akamatsu:2008ge,Das:2010tj,He:2011qa,Alberico:2011zy,Song:2015ykw}. As implied by the large elliptic flow coefficient ($v_2$) of low $p_\mathrm{T}$ heavy flavor mesons that are comparable to light flavor hadrons decayed from the QGP~\cite{STAR:2017kkh,CMS:2022vfn}, these quasi-elastic scatterings should involve strong non-perturbative interactions that rapidly drive heavy quarks towards thermal equilibrium with the QGP. Considerable efforts have been devoted to introducing non-perturbative effects for heavy quarks, such as applying large coupling constant or hard thermal loop propagator to perturbative calculations~\cite{Gossiaux:2008jv,Alberico:2011zy,Uphoff:2012gb,Cao:2016gvr}, incorporating thermal parton masses extracted from the lattice QCD data in quasi-particle models of heavy-light parton scatterings~\cite{Plumari:2011mk,Song:2015ykw,Liu:2021dpm}, or replacing the picture of quasi-particle scattering by heavy quark scattering with a general potential or spectral function inside the QGP~\cite{He:2011qa,Liu:2016ysz,Liu:2017qah,Xing:2021xwc}. With these tools, we are now able to extend studies on heavy quarks from investigating their dynamics inside the QGP to utilizing them to constrain the medium properties, including the transport coefficients~\cite{Xu:2017obm,Cao:2018ews,Ke:2018tsh,Rapp:2018qla,Xu:2018gux,Li:2019lex,Karmakar:2023ity}, the in-medium color force~\cite{Liu:2018syc}, and even the equation of state of the QGP~\cite{Liu:2023rfi}.

A long-standing crucial topic of heavy quarks is the mass and flavor dependence of parton energy loss inside the QGP. Perturbative QCD (pQCD) calculations suggest stronger energy loss for gluons than for quarks due to their different color factors, and weaker energy loss for heavier quarks in both elastic and inelastic processes~\cite{Cao:2017hhk}. This hierarchy is supported by the larger nuclear modification factor ($R_\mathrm{AA}$) and smaller elliptic flow coefficient ($v_2$) of bottom ($b$) decayed electrons than charm ($c$) decayed electrons observed at low $p_\mathrm{T}$~\cite{STAR:2021uzu,Kelsey:2020bms}. On the other hand, at high $p_\mathrm{T}$, current experimental data show comparable $R_\mathrm{AA}$ between charged hadrons, $D$ mesons and $B$ mesons~\cite{CMS:2017qjw}. Through a series of theoretical investigations~\cite{Buzzatti:2011vt,Djordjevic:2013pba,Xing:2019xae}, it has been recognized that this coincidence results from the interplay between the initial spectra, energy loss and fragmentation functions of different parton species. However, the hierarchy of parton energy loss itself has seldom been challenged. Interestingly, using a linear Boltzmann transport model that combines Yukawa and string types of interactions between heavy quarks and the QGP, we will show that the energy loss of bottom quarks is not necessarily always smaller than that of charm quarks. The hierarchy of their energy loss depends on the competition between the string interaction and the dead cone effect, which is further influenced by the heavy quark momentum and the medium temperature. Within this model, we obtain a larger $R_\mathrm{AA}$ and a smaller $v_2$ of bottom mesons (leptons) than charm mesons (leptons) at low $p_\mathrm{T}$ as observed at both RHIC and LHC, but predict an inverse order at high $p_\mathrm{T}$. Furthermore, the stronger string interaction experienced by heavier particles is also manifested in their larger momentum space transport coefficient ($\hat{q}$). We note that in certain kinematic or temperature regions, a smaller $R_\mathrm{AA}$ of $B$ mesons than $D$ mesons was also seen in earlier studies~\cite{Xu:2014ica,Xu:2015bbz}, and a smaller spatial diffusion coefficient $D_\mathrm{s}$ of bottom quarks than charm quarks was shown in other model calculations~\cite{Ke:2018tsh,Das:2016llg,Liu:2016ysz,Sambataro:2023tlv}. However, the origin of these nonintuitive hierarchies have not been clearly identified in literature yet. This will be investigated in detail in our present work.

\section{Heavy quark evolution in relativistic heavy-ion collisions}
\label{sec:model}


To start with, we use the Glauber model to initialize the spatial distributions of both the heavy quark production vertices and the energy density of the QGP. The transverse momentum spectra of the initial heavy quarks are calculated using the fixed-order-next-to-leading-log (FONLL) package~\cite{Cacciari:2001td,Cacciari:2012ny,Cacciari:2015fta}, together with the CT14NLO parton distribution function~\cite{Dulat:2015mca} for free nucleons, which is modified with the EPPS16 parameterization~\cite{Eskola:2016oht} for nucleons bounded inside nuclei. These $p_\mathrm{T}$ spectra are assumed to be rapidity independent around mid-rapidity.

With the initial condition above, we use the (3+1)-D viscous hydrodynamic model CLVisc~\cite{Pang:2018zzo,Wu:2018cpc,Wu:2021fjf} to simulate the evolution of the QGP medium starting from an initial proper time of $\tau_0=0.6$~fm/$c$, and apply the linear Boltzmann transport (LBT) model~\cite{Cao:2016gvr,Xing:2021xwc,Luo:2023nsi} to describe the heavy quark interaction with the QGP. In the LBT model, we use the following Boltzmann equation to evolve the phase space distribution of heavy quarks
\begin{eqnarray}
  \label{eq:boltzmann1}
  p_a \cdot\partial f_a(x_a,p_a)=E_a (\mathcal{C}_\mathrm{el}+\mathcal{C}_\mathrm{inel}),
\end{eqnarray}
where $x_a=(t,\vec{x}_a)$ and $p_a=(E_a,\vec{p}_a)$ are the 4-position and 4-momentum of heavy quarks respectively, and the collision integral on the right hand side includes contributions from both elastic and inelastic scatterings.

For the elastic process, the collision rate can be extracted from the above equation as 
\begin{align}
  \label{eq:gamma_el}
   \Gamma^\mathrm{el}_{ab \rightarrow cd} & (\vec{p}_a, T) = \frac{\gamma_b}{2E_a}\int \frac{d^3 p_b}{(2\pi)^3 2E_b} \frac{d^3 p_c}{(2\pi)^3 2E_c} \frac{d^3 p_d}{(2\pi)^3 2E_d}\nonumber
  \\ \times\,& f_b (\vec{p}_b, T) [1 \pm f_d (\vec{p}_d, T)] \,\theta (s - (m_a + \mu_D)^2) \nonumber
  \\ \times\,& (2\pi)^4 \delta^{(4)}(p_a + p_b - p_c -p_d) |\mathcal{M}_{ab \rightarrow cd}|^2,
\end{align}
in which $T$ is the local temperature of the medium, $|\mathcal{M}_{ab \rightarrow cd}|^2$ is the matrix element of the $ab\rightarrow cd$ scattering, with $b$ representing thermal partons from the medium, $c$ and $d$ representing the final state heavy quarks and thermal partons respectively. The factor $\gamma_b$ denotes the spin-color degeneracy of parton $b$, $\mu_D$ is the Debye screening mass that will be specified later. We use thermal distributions for $f_b$ and $f_d$ above, and use $s,t,u$ for the Mandelstam variables. The thermal light partons are assumed massless in this work, and the bare quark mass 1.27~GeV is used for charm quarks and 4.19~GeV for bottom quarks.

We follow our previous study~\cite{Xing:2021xwc} to include both Yukawa and string interactions between heavy quarks and the QGP. A Cornell-type potential is assumed between a heavy quark and a thermal light quark:
\begin{align}
  V(r)=V_\mathrm{Y}(r)+V_\mathrm{S}(r)=-\frac{4}{3}\alpha_\mathrm{s}\frac{e^{-m_d r}}{r}-\frac{\sigma e^{-m_s r}}{m_\mathrm{s}},
\label{eq:potential}
\end{align}
which includes both a short-range Yukawa (Y) term and a long-range string (S) term, with $\alpha_\mathrm{s}$ and $\sigma$ being their coupling strengths respectively. In this model, we use the Yukawa term to approximate the perturbative interaction, and use the string term to approximate the non-perturbative interaction. The temperature dependent screening masses for the two terms are parametrized as 
$m_d=a+bT$ and $m_s=\sqrt{a_s+b_s T}$. Here, we use the model parameters listed in Tab.~\ref{table:parameters}, which are determined in Ref.~\cite{Xing:2021xwc} based on the $D$ meson observables at RHIC and LHC. Note that the potential we extract from the open heavy flavor data is not necessarily the same with that between a static quark-antiquark pair from the lattice QCD calculation~\cite{Burnier:2014ssa}. However, with the parameters here, the potential given by Eq.~(\ref{eq:potential}) appears similar to the lattice data~\cite{Xing:2021xwc}. The value of $m_d$ in the Yukawa term is also applied for the screening mass $\mu_D$ in Eq.~(\ref{eq:gamma_el}).

\begin{table}[h]
\centering
{\renewcommand{\arraystretch}{1.2}
\begin{tabular}{|c|c|c|c|c|c|}
\hline
$\alpha_\mathrm{s}$ &\, $\sigma$ (GeV$^2$) \,& \,$a$ (GeV)\, & $b$ & \, $a_s$ (GeV$^2$) \, &\, $b_s$ (GeV)\,  \\ \hline
\,0.27\,   & 0.45  & 0.20 & \,2.0\, & 0  & 0.10  \\ \hline
\end{tabular}}
\caption{Parameters of the interaction potential between a heavy quark and a medium parton.}
\label{table:parameters}
\end{table}

We can then perform Fourier transformation on Eq.~(\ref{eq:potential}) into the momentum space as 
\begin{eqnarray}
  \label{eq:potential_q}
  V(\vec{q}) = - \frac{4\pi \alpha_\mathrm{s} C_F}{m_d^2+|\vec{q}|^2} - \frac{8\pi \sigma}{(m_s^2+|\vec{q}|^2)^2},
\end{eqnarray}
with $C_F=4/3$ being the color factor and $\vec{q}$ being the momentum exchange between a heavy quark and a medium parton. This momentum space potential is further used as an effective gluon propagator in evaluating the scattering matrix as
\begin{align}
  \label{eq:Matrix_cq}
 i \mathcal{M}  =\,& i\mathcal{M_\mathrm{Y}} + i\mathcal{M_\mathrm{S}}\nonumber
 \\=\,& \overline{u}(p') \gamma^{\mu} u(p) V_\mathrm{Y}(\vec{q}) \overline{u}(k') \gamma_{\mu} u(k)\nonumber
 \\&+ \overline{u}(p') u(p) V_\mathrm{S}(\vec{q}) \overline{u}(k') u(k),
\end{align}
in which we keep the vector interaction vertex for the Yukawa term for consistency with the perturbative calculation of a two-body scattering at the leading order (LO), and assume a scalar interaction vertex for the string term~\cite{Riek:2010fk}. 

By setting $|\vec{q}|^2=-t$, and summing (averaging) over the final (initial) state spin degrees of freedom, we obtain the matrix element square of the $Qq\rightarrow Qq$ process as 
\begin{align}
  \label{eq:M2_cq}
|\mathcal{M}_{Qq}|^2 &= \frac{64\pi^2 \alpha_\mathrm{s}^2}{9} \frac{(s-m_Q^2)^2+(m_Q^2-u)^2+2m_Q^2 t}{(t-m_d^2)^2}\nonumber\\
 &+ \frac{(8\pi \sigma)^2}{N_c^2 -1} \frac{t^2-4m_Q^2 t}{(t-m_s^2)^4}.
\end{align}
On the right hand side of the equation above, the first term comes from the Yukawa interaction while the second from the string interaction. There is no interference between $\mathcal{M_\mathrm{Y}}$ and $\mathcal{M_\mathrm{S}}$ due to their different types of interaction vertices. We introduce an additional $1/(N_c^2-1)$ factor for the gluon field here in order to reproduce the color factor $C_F^2/(N_c^2-1)$ in the well established LO result of heavy-light quark scattering~\cite{Combridge:1978kx} in the first term. 

Based on the form of $|\mathcal{M}_{Qq}|^2$ above, one may write the matrix element square of the $Qg\rightarrow Qg$ process as
\begin{align}
  \label{eq:M2_cg}
  |\mathcal{M}_{Qg}&|^2 =  \frac{64\pi^2 \alpha_\mathrm{s}^2}{9} \frac{(s-m_Q^2)(m_Q^2-u)+2m_Q^2 (s+m_Q^2)}{(s-m_Q^2)^2}\nonumber
  \\+\, & \frac{64\pi^2 \alpha_\mathrm{s}^2}{9} \frac{(s-m_Q^2)(m_Q^2-u)+2m_Q^2 (u+m_Q^2)}{(u-m_Q^2)^2}\nonumber
  \\+\, & 8\pi^2 \alpha_\mathrm{s}^2 \frac{5m_Q^4 + 3m_Q^2t -10m_Q^2u + 4t^2 + 5tu + 5u^2}{(t-m_d^2)^2}\nonumber
  \\+\, & 8\pi^2 \alpha_\mathrm{s}^2 \frac{(m_Q^2-s)(m_Q^2-u)}{(t-m_d^2)^2}\nonumber
  \\+\, & 16\pi^2 \alpha_\mathrm{s}^2 \frac{3m_Q^4 -3m_Q^2s - m_Q^2u + s^2}{(s-m_Q^2)(t-m_d^2)}\nonumber
  \\+\, & \frac{16\pi^2 \alpha_\mathrm{s}^2}{9} \frac{m_Q^2 (4m_Q^2 - t)}{(s - m_Q^2)(m_Q^2 - u)}\nonumber
  \\+\, & 16\pi^2 \alpha_\mathrm{s}^2 \frac{3m_Q^4 - m_Q^2s -3m_Q^2u +u^2}{(t - m_d^2)(u - m_Q^2)}\nonumber
  \\+\, & \frac{C_A}{C_F} \frac{(8\pi \sigma)^2}{N_c^2 -1} \frac{t^2-4m_Q^2 t}{(t-m_s^2)^4},
\end{align}
in which the last term is for the string interaction which differs from that in Eq.~(\ref{eq:M2_cq}) by the color $C_A/C_F$, and the other terms are from the LO perturbative calculation~\cite{Combridge:1978kx} which takes into account $s$, $t$ and $u$-channel scatterings between a heavy quark and a gluon. Note that for the $\alpha_\mathrm{s}^2$ factors in the perturbative parts of Eqs.(~\ref{eq:M2_cq}) and~(\ref{eq:M2_cg}), one $\alpha_\mathrm{s}$ is from the vertex connecting the exchanged gluon with the thermal parton, and the other is from the vertex connecting the heavy quark and the exchanged gluon. We use the fixed value $\alpha_\mathrm{s}=0.27$ in Tab.~\ref{table:parameters} for the former, while the latter is assumed to run with the heavy quark energy and the medium temperature as $\alpha_\mathrm{s} = 4\pi / [9 \mathrm{ln}(2ET/\Lambda^2)]$~\cite{Cao:2017hhk} with $\Lambda=0.2$~GeV.

For the inelastic scattering process, we connect the scattering rate to the average number of medium-induced gluons per unit time\footnote{We use the notation $\bar{t}$ for time here to distinguish from the Mandelstam variable $t$.} as
\begin{eqnarray}
  \label{eq:gamma_inel}
  \Gamma_\mathrm{inel}^a (E_a, T, \bar{t}) = \int dxdl_{\bot}^2 \frac{dN_g^a}{dxdl_\bot ^2 d\bar{t}},
\end{eqnarray}
where the emitted gluon spectrum is taken from the higher-twist energy loss calculation~\cite{Wang:2001ifa,Zhang:2003wk,Majumder:2009ge},
\begin{eqnarray}
  \label{eq:gluon_spectrum}
  \frac{dN_g^a}{dxdl_\bot ^2 d\bar{t}} = \frac{2C_A \alpha_\mathrm{s} P_a(x) l_{\bot}^4 \hat{q}_a}{\pi (l_{\bot}^2 + x^2 m_a^2)^4} \sin^2 \left(\frac{\bar{t}-\bar{t}_i}{2\tau_f}\right).
\end{eqnarray}
Here, $x$ and $l_\perp$ denote the fractional energy and the transverse momentum of the emitted gluon with respect to its parent heavy quark, $P_a(x)$ is the $Q\rightarrow Qg$ splitting function, $\bar{t}_i$ is the production time of the heavy quark (or the time of the previous splitting), $\tau_f = 2E_a x(1-x)/(l_{\bot}^2+x^2m_a^2)$ is the gluon formation time, and the running coupling $\alpha_\mathrm{s} = 4\pi / [9 \mathrm{ln}(2ET/\Lambda^2)]$ is used. The jet transport coefficient $\hat{q}_a$ quantifies the transverse momentum broadening square of the jet parton per unit time -- $d\langle k_\perp^2\rangle/d\bar{t}$ -- due to elastic scatterings, which can be evaluated from Eq.~(\ref{eq:gamma_el}) with an additional weight factor $k^2_\perp=[\vec{p}_c-(\vec{p}_c\cdot \hat{p}_a)\hat{p}_a]^2$ inside the integral.

Using Eqs.~(\ref{eq:gamma_el}) and~(\ref{eq:gamma_inel}), one may construct the scattering probability during a time interval $\Delta\bar{t}$ as $P_\mathrm{el/inel}^a = 1-e^{-\Gamma_\mathrm{el/inel}^a \Delta \bar{t}}$ for elastic and inelastic processes separately, where $\Gamma_\mathrm{el}^a=\sum_{bcd}\Gamma^\mathrm{el}_{ab\rightarrow cd}$ sums over all possible elastic scattering channels for parton $a$. The total scattering probability is then given by
\begin{eqnarray}
  \label{eq:prob_total}
  P_\mathrm{tot}^a = 1 - e^{-(\Gamma^a_\mathrm{el}+\Gamma^a_\mathrm{inel}) \Delta \bar{t}} = P_\mathrm{el}^a + P_\mathrm{inel}^a - P_\mathrm{el}^a P_\mathrm{inel}^a,
\end{eqnarray}
which can be understood as the sum of pure elastic process without inducing gluon emission $P_\mathrm{el}^a(1-P_\mathrm{inel}^a)$ and inelastic process $P_\mathrm{inel}^a$. With these probabilities and differential rates, we implement Monte-Carlo simulation of heavy quark scatterings with the QGP. At a given time step, we first boost each heavy quark into the local rest frame of the hydrodynamic medium. Using the local temperature of the medium, we update the 4-momentum of the heavy quark according to the Boltzmann approach and then boost it back to the global frame and propagate it to the location of the next time step.

\begin{figure}[tbp!]
  \centering
  \includegraphics[width=0.87\linewidth]{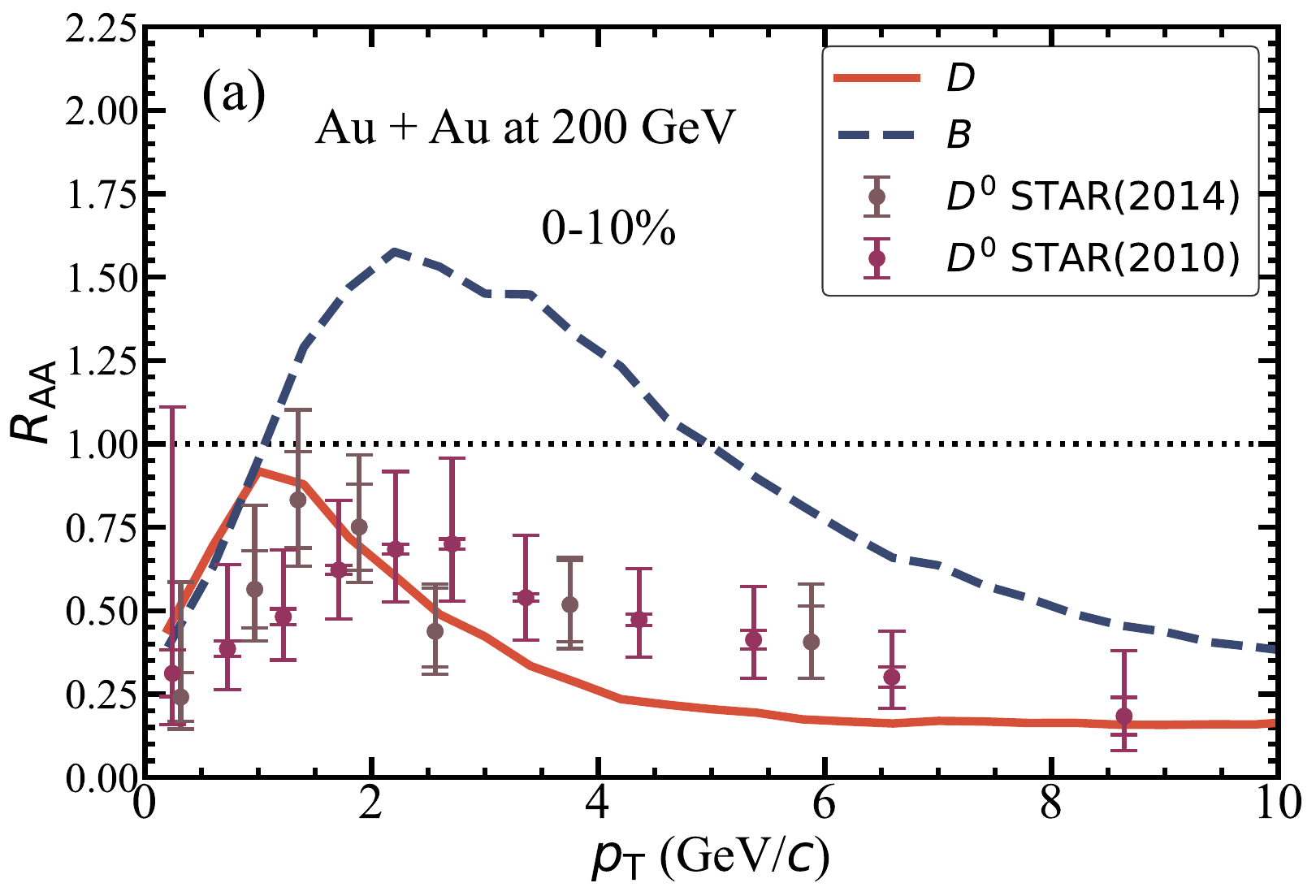}
  \includegraphics[width=0.87\linewidth]{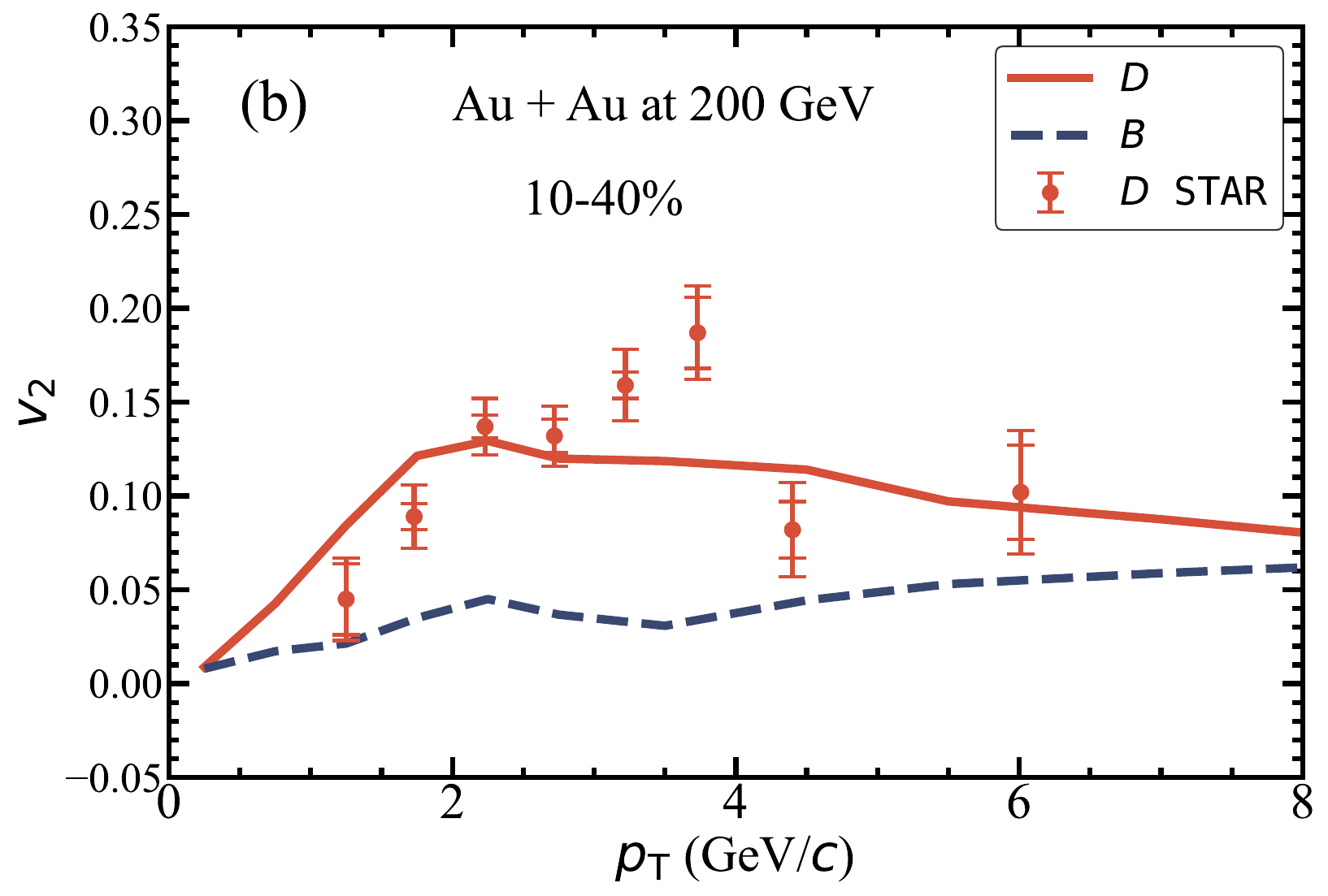}
  \caption{(Color online) The $R_\mathrm{AA}$ (upper panel) and $v_2$ (lower panel) of heavy mesons in Au+Au collisions at $\sqrt{s_{\mathrm{NN}}}=200\ \mathrm{GeV}$, compared between $D$ meson, $B$ meson and the available data from STAR~\cite{STAR:2014wif,STAR:2018zdy,STAR:2017kkh}.}
  \label{fig:meson}
\end{figure}

\begin{figure}[tbp!]
  \centering
  \includegraphics[width=0.87\linewidth]{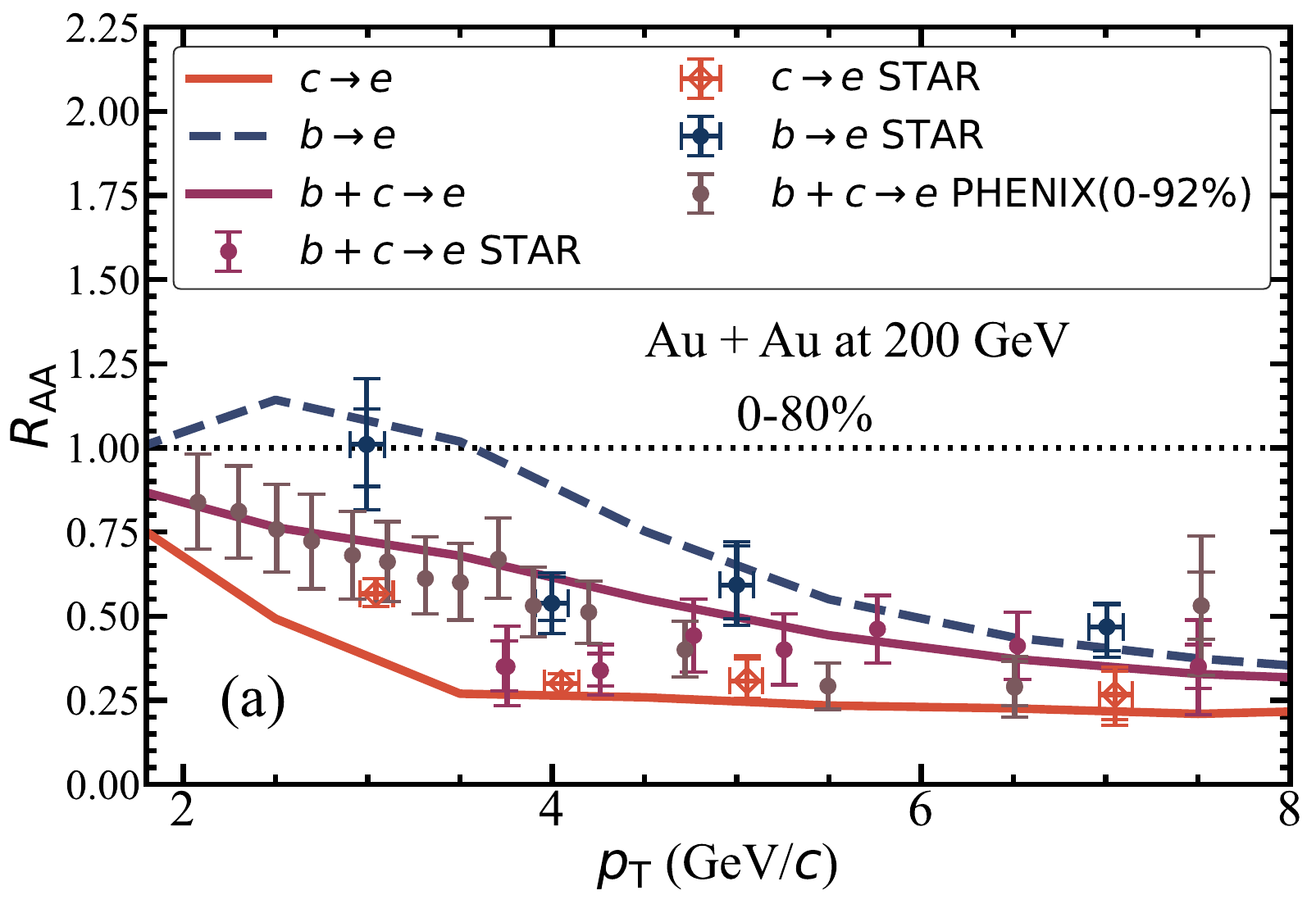}
  \includegraphics[width=0.87\linewidth]{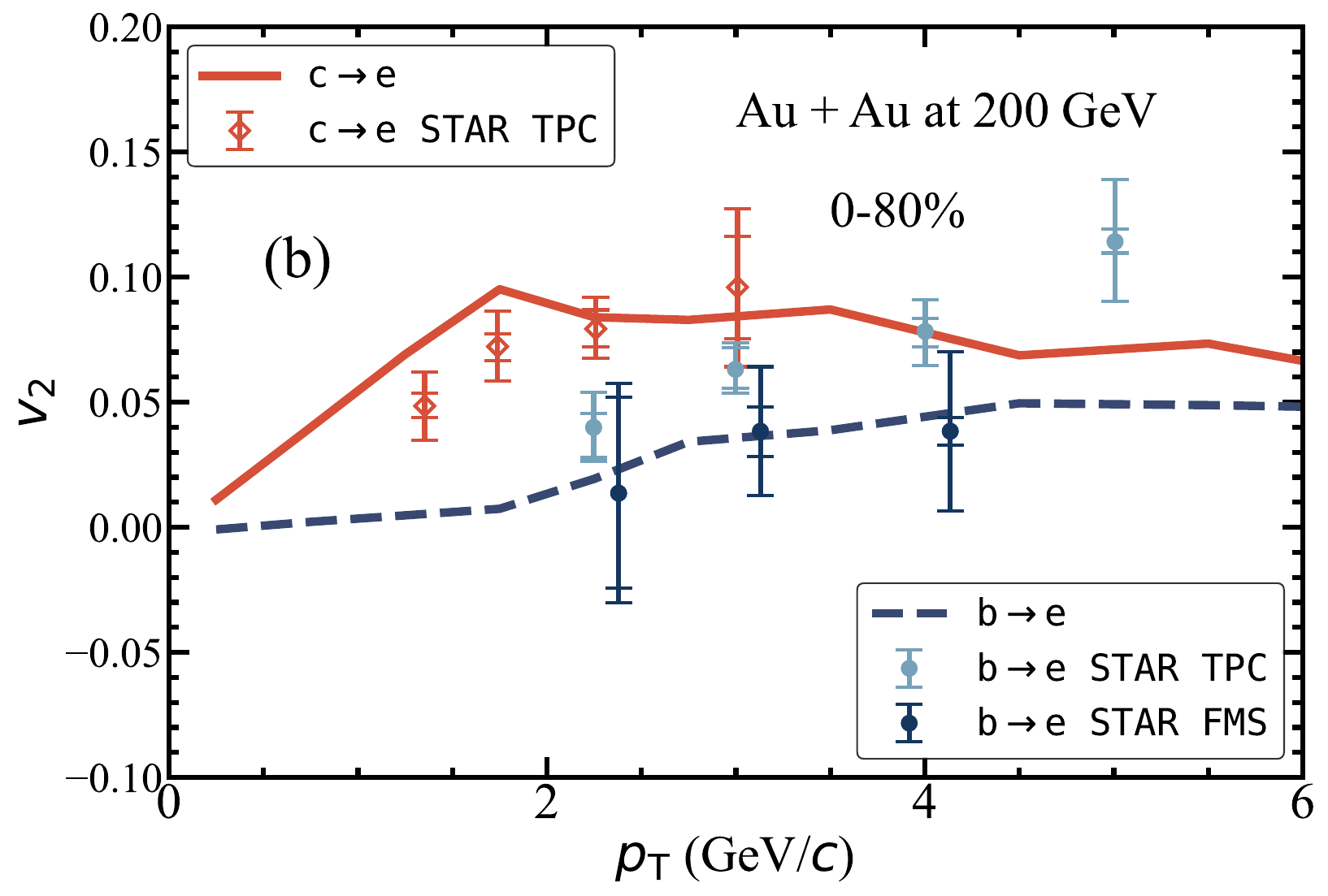}
  \caption{(Color online) The $R_\mathrm{AA}$ (upper panel) and $v_2$ (lower panel) of heavy flavor decayed electrons in Au+Au collisions at $\sqrt{s_{\mathrm{NN}}}=200\ \mathrm{GeV}$, compared between $c$-decayed electrons, $b$-decayed electrons, their mixture and the corresponding data at RHIC~\cite{PHENIX:2010xji,STAR:2021uzu,Kelsey:2020bms}. The $R_\mathrm{AA}$ data of mixed electrons in other centralities can also be found in Refs.~\cite{STAR:2006btx,STAR:2023qfk}.}
  \label{fig:electron}
\end{figure}

Upon exiting the QGP medium (the hypersurface of a temperature at $T_\mathrm{c}=165$~MeV in this work), heavy quarks are converted into heavy flavor hadrons via our hybrid fragmentation-coalescence model developed in Ref.~\cite{Cao:2019iqs}. The momentum dependent coalescence probability is given by the wavefunction overlap between the free quark state and the hadron state. We assume simple harmonic oscillator potential between quarks inside a hadron and include both $s$ and $p$-wave states which cover nearly all heavy flavor hadrons listed in the Particle Data Group~\cite{ParticleDataGroup:2022pth}. The only model parameter for hadronization is the oscillator frequency, which is set as $\omega_c=0.24$~GeV for charm hadrons and $\omega_b=0.14$~GeV for bottom hadrons. They are determined by requiring the total coalescence probability for a zero momentum heavy quark to be one. Based on these probabilities, heavy quarks that do not combine with thermal partons from the QGP are converted to hadrons via the Pythia fragmentation~\cite{Sjostrand:2006za}. This hadronization model is able to provide a good description of both charm and bottom hadron chemistry observed at RHIC and LHC~\cite{Cao:2019iqs,CMS:2021mzx}. In the end, we also use Pythia to decay heavy flavor hadrons into leptons. Note that contributions from heavy flavor baryons (e.g. $\Lambda_c$ and $\Lambda_b$) are included in this work. If they are not taken into account, one would obtain larger $R_\mathrm{AA}$'s of heavy flavor mesons and their decayed leptons at low $p_\mathrm{T}$ compared to our results, because of the larger baryon-to-meson ratio in nucleus-nucleus (AA) than in proton-proton (pp) collisions due to the coalescence process~\cite{Cao:2019iqs}. On the other hand, this should not affect the heavy meson (lepton) $R_\mathrm{AA}$ at high $p_\mathrm{T}$, where hadronization is dominated by fragmentation.

\section{Nuclear modification of heavy mesons and their decayed electrons}
\label{sec:Nuclear_modification}

We focus on the two typical observables -- nuclear modification factor ($R_\mathrm{AA}$) and elliptic flow coefficient ($v_2$) of heavy flavor particles in high-energy nuclear collisions. The former measures the ratio between their spectra in AA and pp collisions:
\begin{align}
  R_{\mathrm{AA}}(p_\mathrm{T})\equiv\frac{\mathrm{d}N^{\mathrm{AA}}/\mathrm{d}p_\mathrm{T}}{\mathrm{d}N^{\mathrm{pp}}/\mathrm{d}p_\mathrm{T}\times 	\left \langle N_\mathrm{coll}^{\mathrm{AA}}\right \rangle},
\end{align}
where $\left \langle N_\mathrm{coll}^\mathrm{AA}\right \rangle$ is the average number of nucleon-nucleon binary collisions in each AA collision. The latter quantifies the momentum space anisotropy of the particle production:
\begin{align}
v_2(p_\mathrm{T})\equiv\left \langle\cos(2\phi) \right \rangle=\left \langle \frac{p_x^2-p_y^2}{p_x^2+p_y^2} \right \rangle,
\end{align}
where $\phi$ represents the azimuthal angle of heavy flavor particles in the transverse plane and the average is performed both over particles within each event and across different events. In this work, we use smooth hydrodynamic medium for the QGP and align its second order event plane with the $\hat{x}$ direction in our computational frame. Effect of event-by-event fluctuations on the $R_\mathrm{AA}$ and $v_2$ of hard probe particles was shown small in our earlier studies~\cite{Cao:2017umt,He:2022evt}.

\begin{figure}[tbp!]
  \centering
  \includegraphics[width=0.87\linewidth]{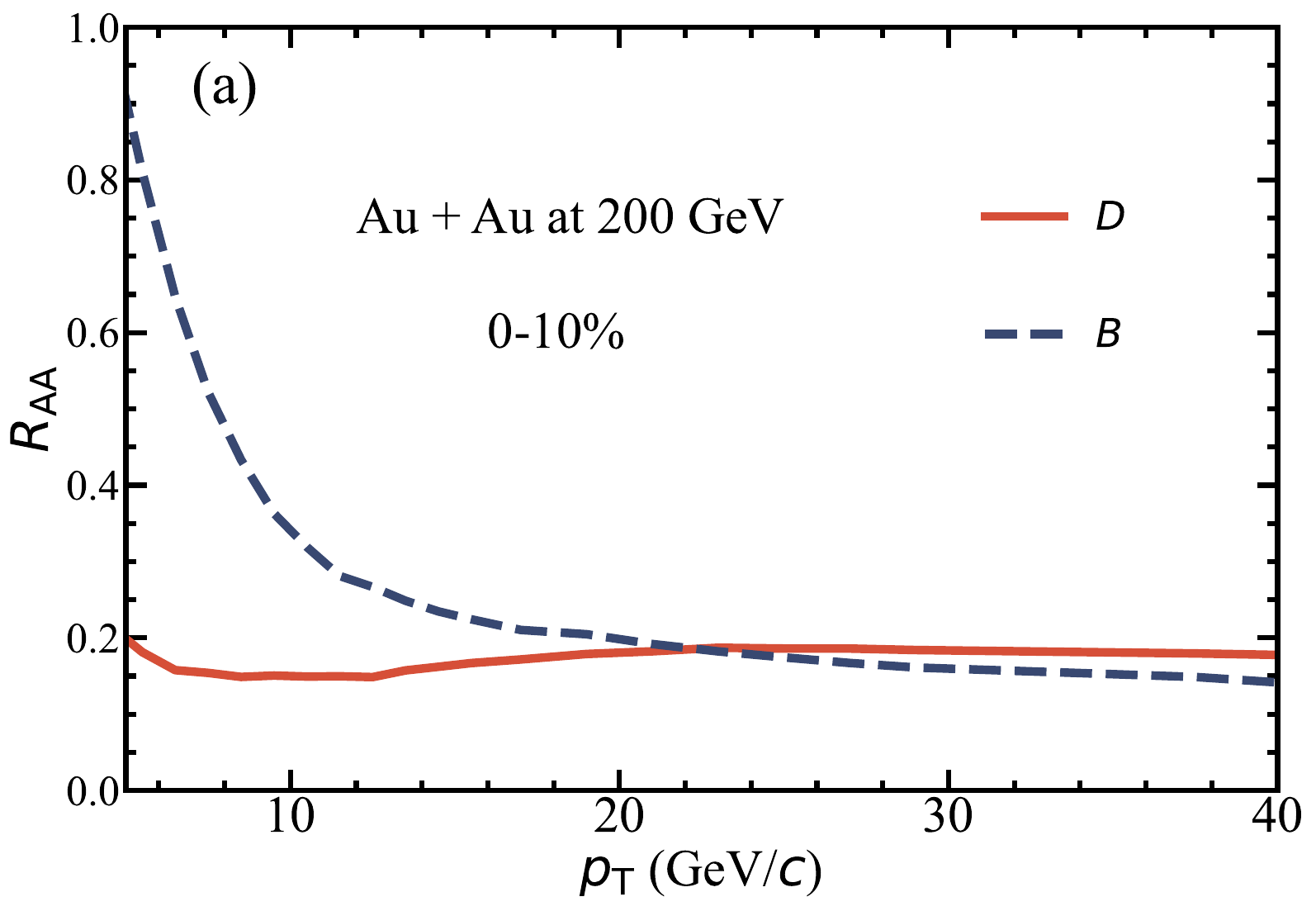}
  \includegraphics[width=0.87\linewidth]{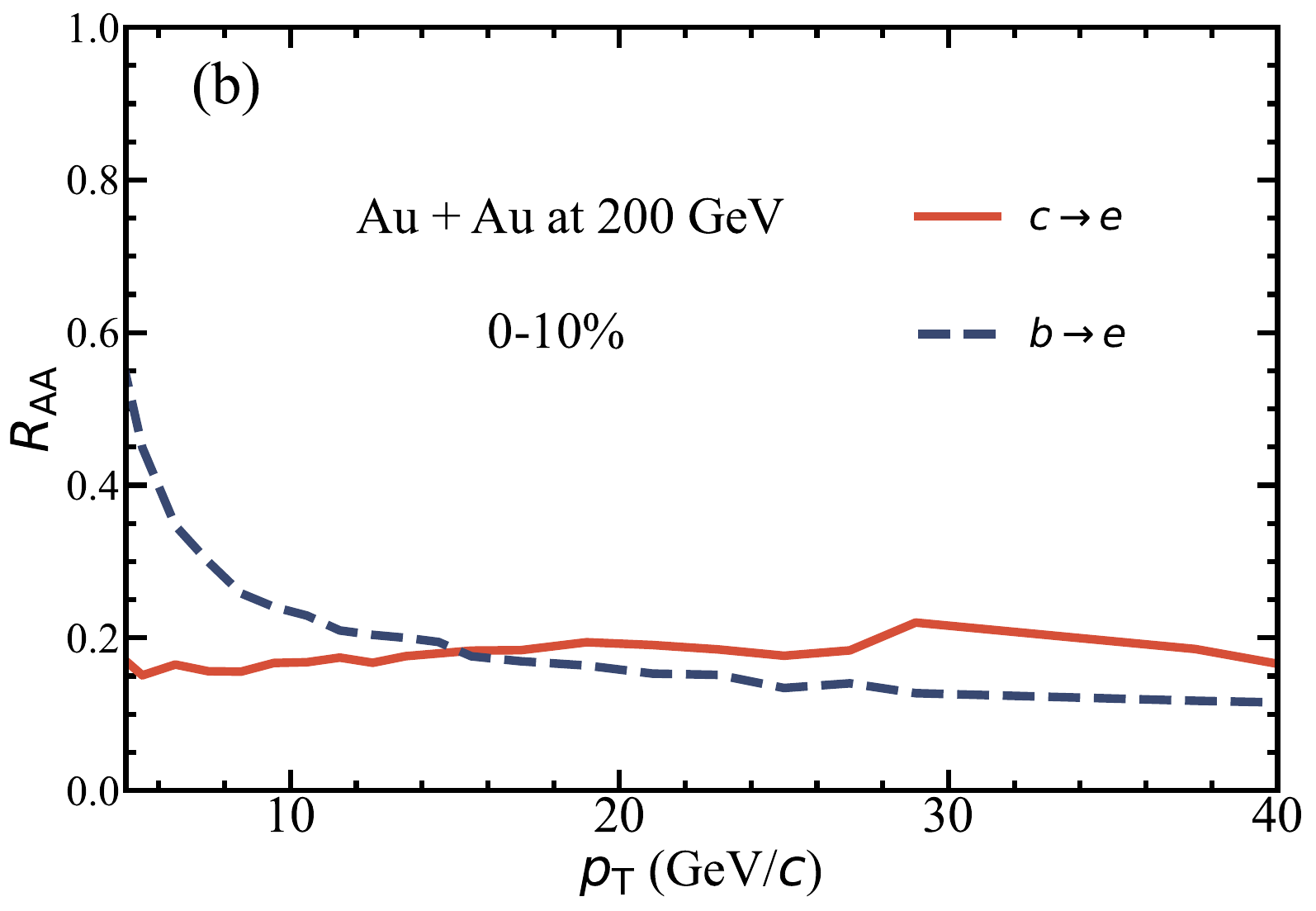}
  \caption{(Color online) The $R_{\mathrm{AA}}$ of heavy flavor mesons (upper panel) and their decayed electrons (lower panel) in Au+Au collisions at $\sqrt{s_{\mathrm{NN}}}=200\ \mathrm{GeV}$, extended to the high $p_\mathrm{T}$ region.}
  \label{fig:RAALO}
\end{figure}

\begin{figure}[tbp!]
  \centering
  \includegraphics[width=0.87\linewidth]{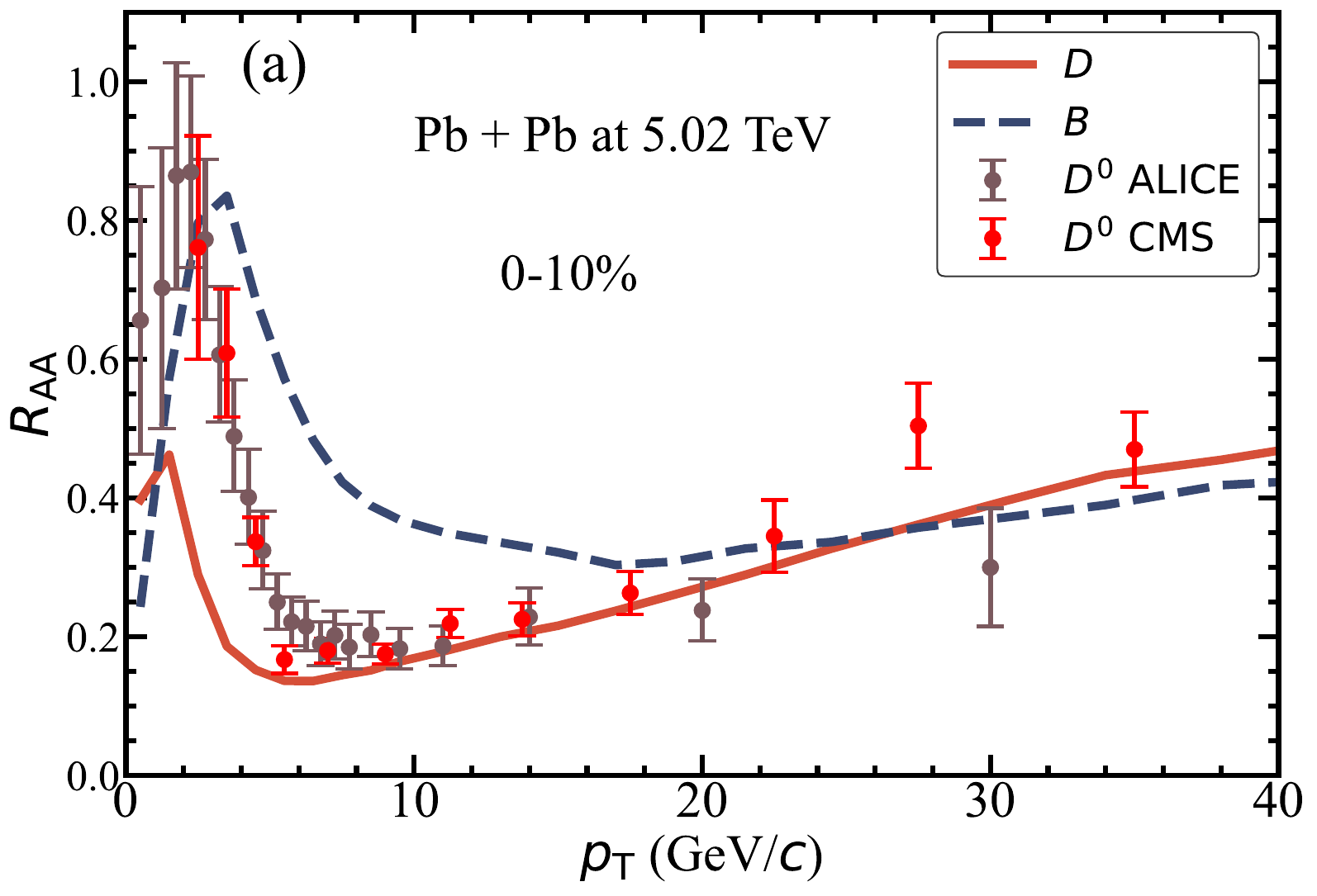}
  \includegraphics[width=0.87\linewidth]{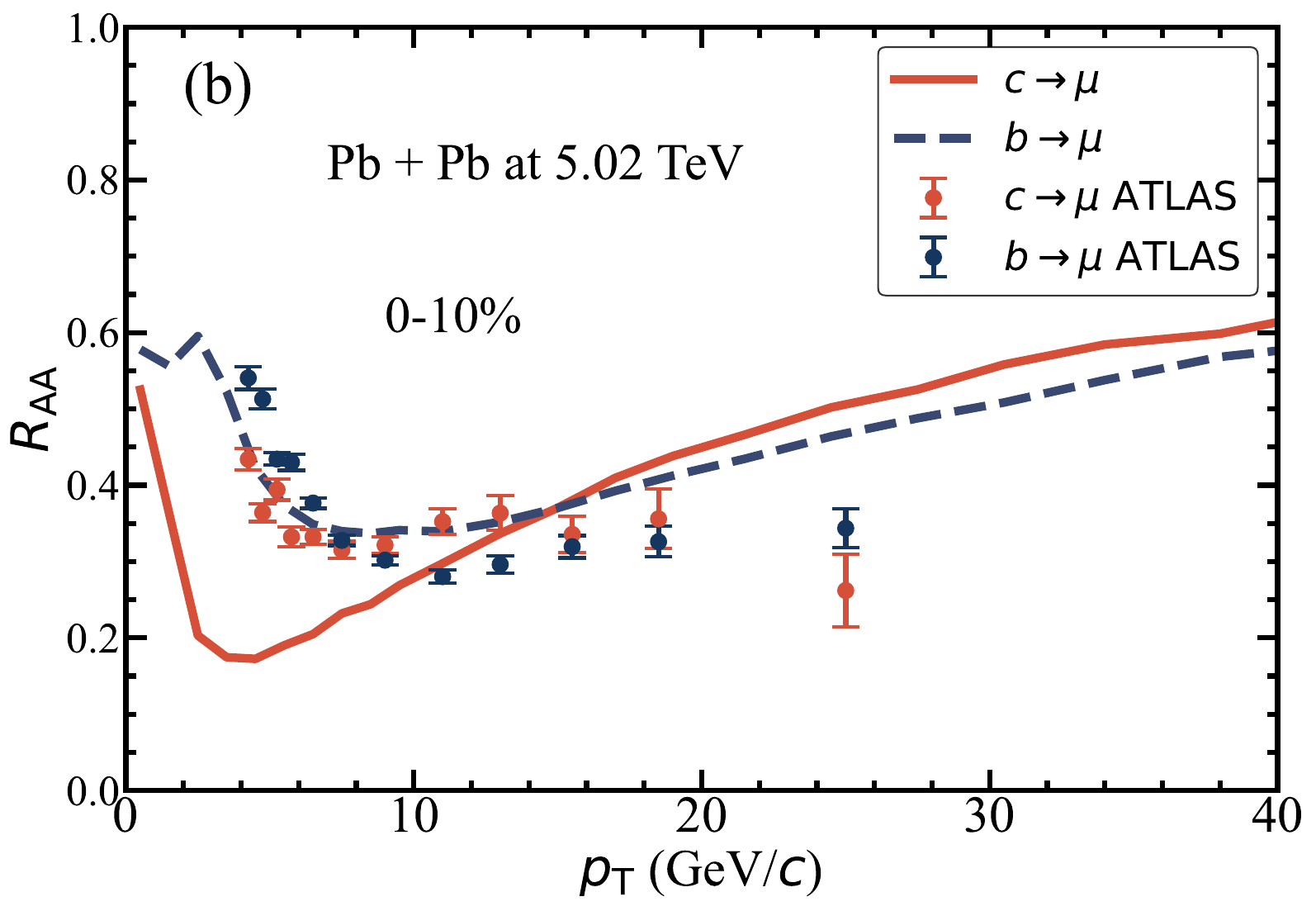}
  \caption{(Color online) The $R_{\mathrm{AA}}$ of heavy flavor mesons (upper panel) and their decayed muons (lower panel) in Pb+Pb collisions at $\sqrt{s_{\mathrm{NN}}}=5.02\ \mathrm{TeV}$. The experimental data are take from ALICE, CMS and ATLAS~\cite{ALICE:2021rxa,CMS:2017qjw,ATLAS:2021xtw}.}
  \label{fig:RAA-LHC}
\end{figure}

We first present the nuclear modification of heavy mesons in Au+Au collisions at $\sqrt{s_{\mathrm{NN}}}=200\ \mathrm{GeV}$ in Fig.~\ref{fig:meson}, compared between $D$ and $B$ mesons. One observes larger $R_\mathrm{AA}$ (upper panel) and smaller $v_2$ (lower panel) of $B$ mesons than $D$ mesons within the $p_\mathrm{T}$ range covered by the current RHIC experiment, consistent with our expectation of weaker energy loss and thus a slower thermalization process of $b$ quarks than $c$ quarks inside the QGP due to the heavier mass of the former. Our results on $D$ mesons are consistent with the available data from the STAR collaboration~\cite{STAR:2014wif,STAR:2018zdy,STAR:2017kkh}.

The mass hierarchy of heavy quark energy loss can be further verified by the nuclear modification of their decayed electrons, as shown in Fig.~\ref{fig:electron}. In the upper panel, we see a larger $R_\mathrm{AA}$ of $b$-decayed electrons than $c$-decayed electrons from both our model calculation and the experimental data, while their mixture is in between. In the lower panel, a smaller $v_2$ is seen for $b$-decayed electrons than $c$-decayed electrons. Our LBT model including both Yukawa and string interactions provides a simultaneous description of the charm and bottom flavor data. The only difference between charm and bottom quarks through our calculation is their masses.

In Fig.~\ref{fig:RAALO}, we extend our calculation to higher $p_\mathrm{T}$ that is beyond the current RHIC measurement but will be covered by the upcoming sPHENIX data. Interestingly, we observe a crossing of $R_\mathrm{AA}$ between $D$ and $B$ mesons around $p_\mathrm{T}\approx 20$~GeV/$c$ in the upper panel. Such crossing also exists in the lower panel between $c$-decayed and $b$-decayed electrons, and is shifted towards lower $p_\mathrm{T}$ during the decay process. The similar crossing patterns can also be seen from our calculation in Fig.~\ref{fig:RAA-LHC} for heavy flavor mesons (upper panel) and their decayed muons (lower panel) in Pb+Pb collisions at $\sqrt{s_\mathrm{NN}}=5.02$~TeV. We have verified that the slightly stronger suppression of $b$-quarks than $c$-quarks at high $p_\mathrm{T}$ is not due to their different initial spectra or hadronization, but is from stronger energy loss of $b$-quarks than $c$-quarks. This is contradictory to one's expectation of the mass hierarchy of parton energy loss and was not seen in our earlier LBT calculations that only includes perturbative interactions between heavy quarks and the QGP. We note that a similar crossing pattern was also seen in the CUJET calculation~\cite{Xu:2014ica,Xu:2015bbz}, although its origin was not clearly identified. In the next section, we will conduct a detailed exploration on this inverse hierarchy of charm {\it vs.} bottom quark energy loss.

\section{Mass dependence of quark energy loss and transport coefficients}
\label{sec:EandDs}

\begin{figure}[tbp!]
  \centering
   \includegraphics[width=0.84\linewidth]{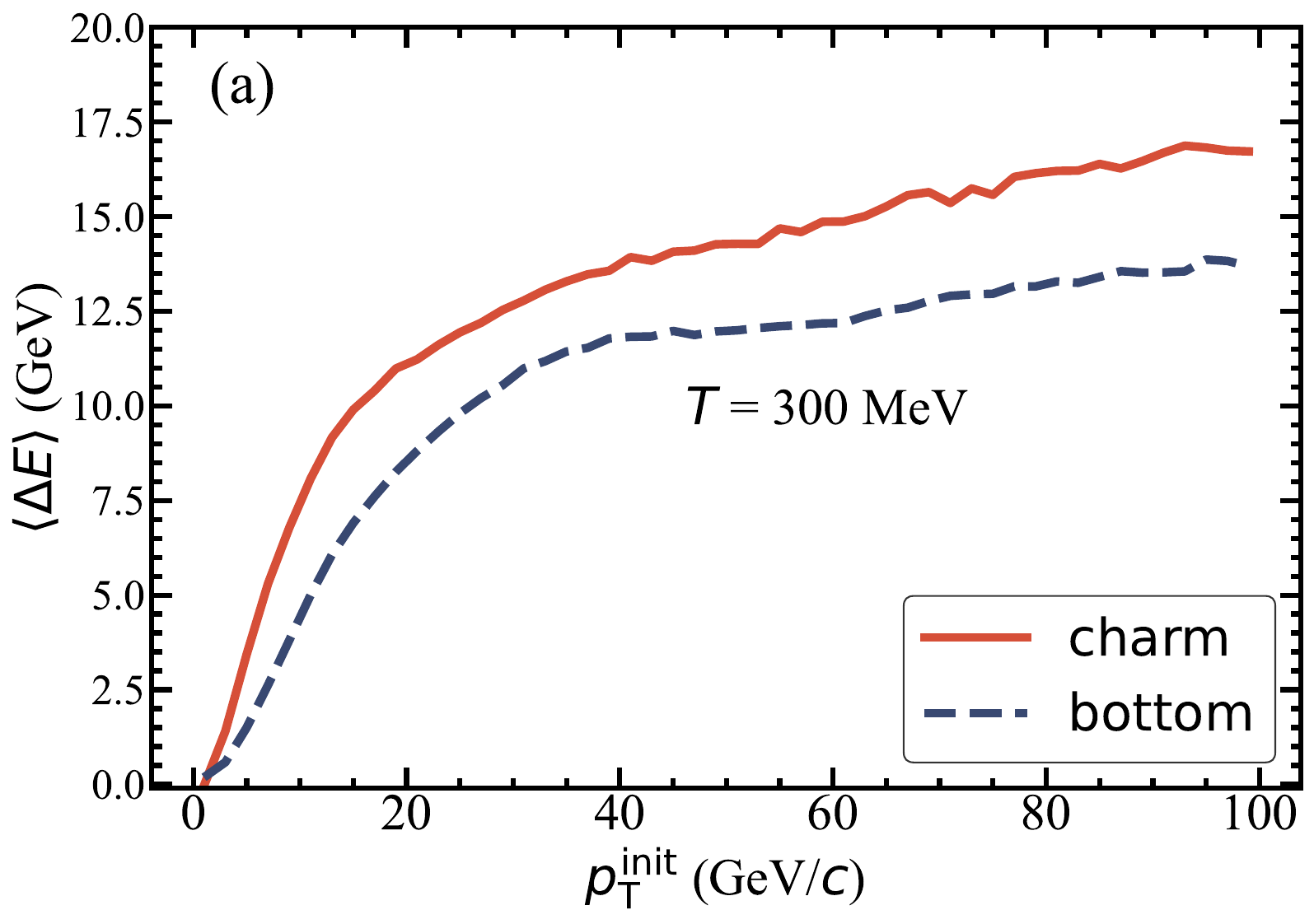}
    \includegraphics[width=0.84\linewidth]{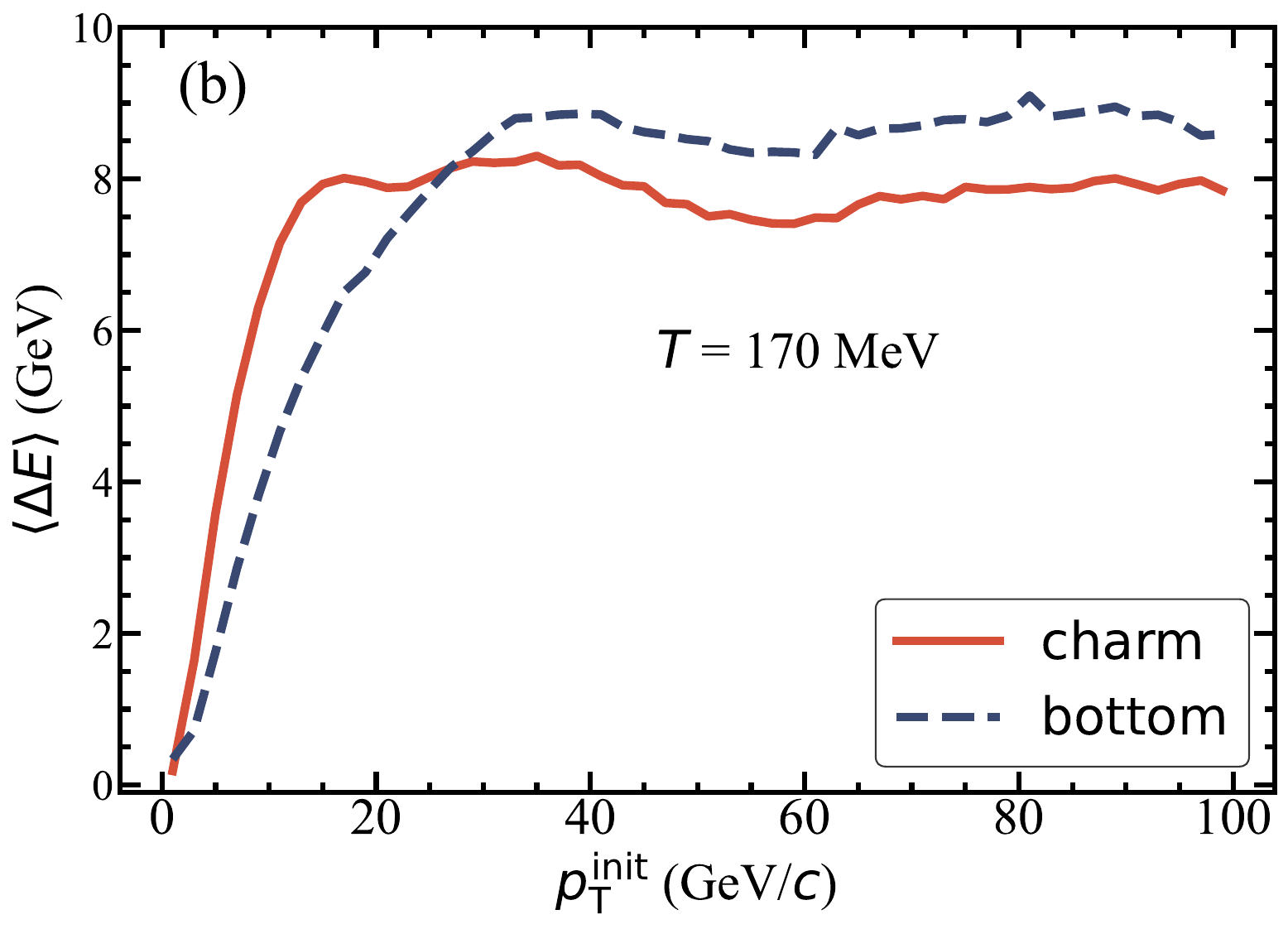}
  \caption{(Color online) The average energy loss of charm {\it vs.} bottom quarks at $t=6$~fm/$c$ through static media at (a) $T=300$~MeV and (b) $T=170$~MeV, as functions of their initial transverse momenta.}
  \label{fig:energyloss}
\end{figure}

To understand the crossing of $R_\mathrm{AA}$ between charm and bottom mesons (electrons), we first study the energy loss of heavy quarks through a static medium with a given temperature. At a time of $t=6$~fm/$c$, we present the average energy loss of charm {\it vs.} bottom quarks as a function of their initial transverse momenta ($p_\mathrm{T}^\mathrm{init}$) in Fig.~\ref{fig:energyloss}. In the upper panel, we use a relatively high temperature ($T=300$~MeV) and observe charm quarks lose more energy than bottom quarks. To the contrary, the opposite order can be seen in the lower panel above $p_\mathrm{T}^\mathrm{init}\approx 27$~GeV when the medium temperature is low ($T=170$~MeV). This indicates the stronger string interaction at lower temperature, especially near the phase transition temperature $T_\mathrm{c}$, could lead to stronger energy loss of bottom quarks than charm quarks.

\begin{figure}[tbp!]
  \centering
  \includegraphics[width=0.87\linewidth]{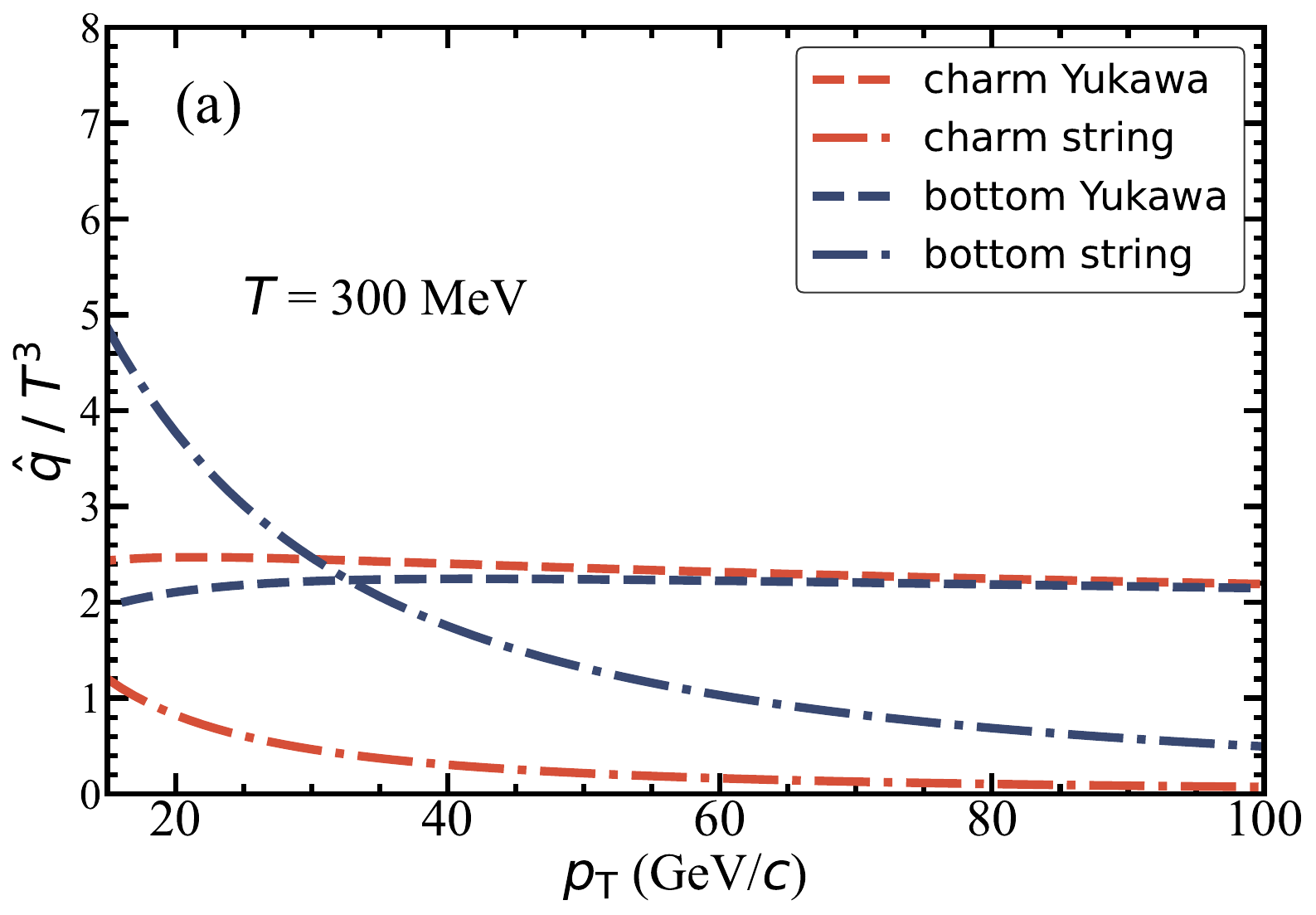}
  \includegraphics[width=0.87\linewidth]{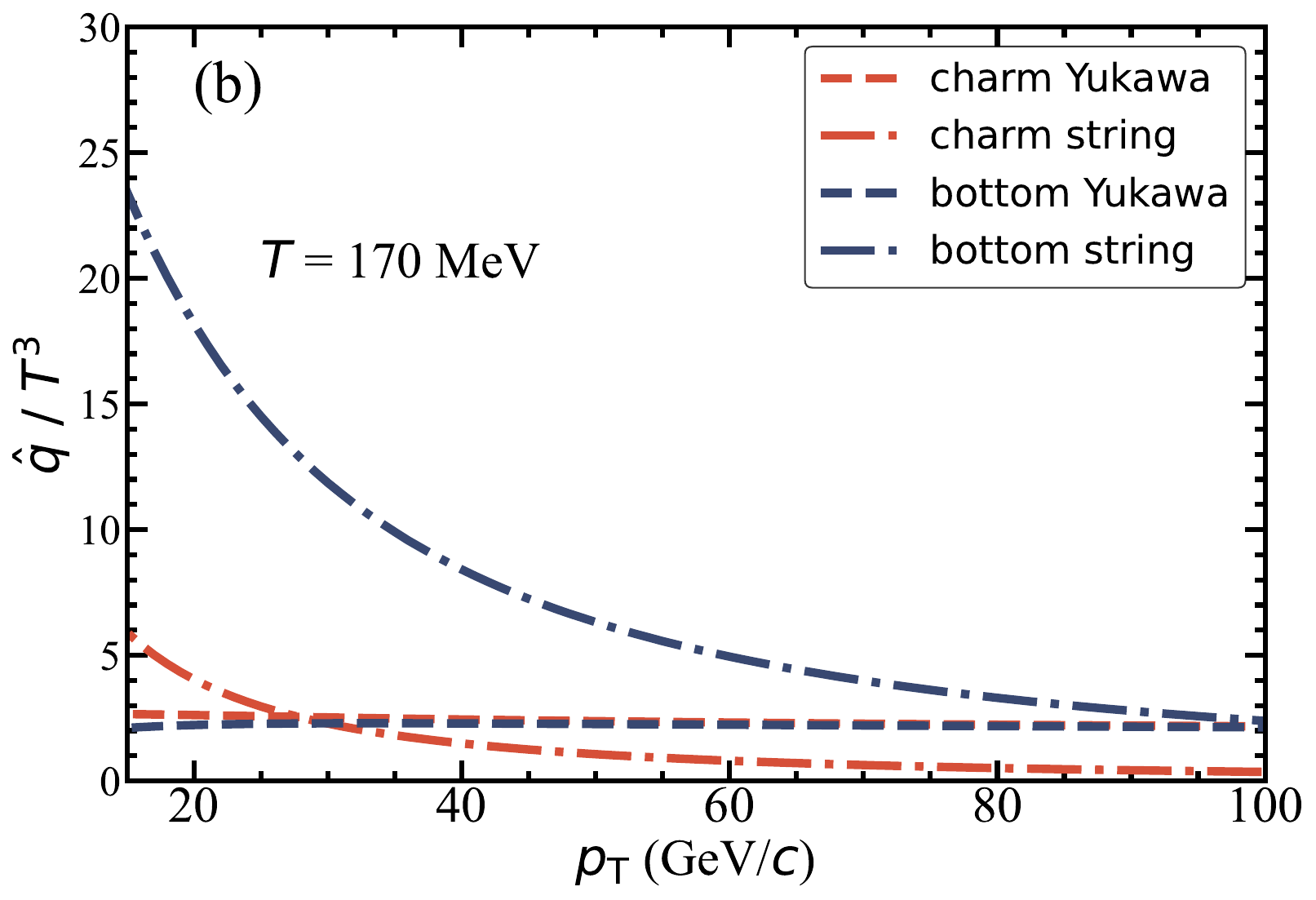}
  \caption{(Color online) Momentum dependence of jet transport coefficient at (a) $T=300$~MeV and (b) $T=170$~MeV, compared between Yukawa and string contributions to heavy-quark-QGP interactions.}
  \label{fig:qhatp}
\end{figure}

\begin{figure}[tbp!]
  \centering
  \includegraphics[width=0.87\linewidth]{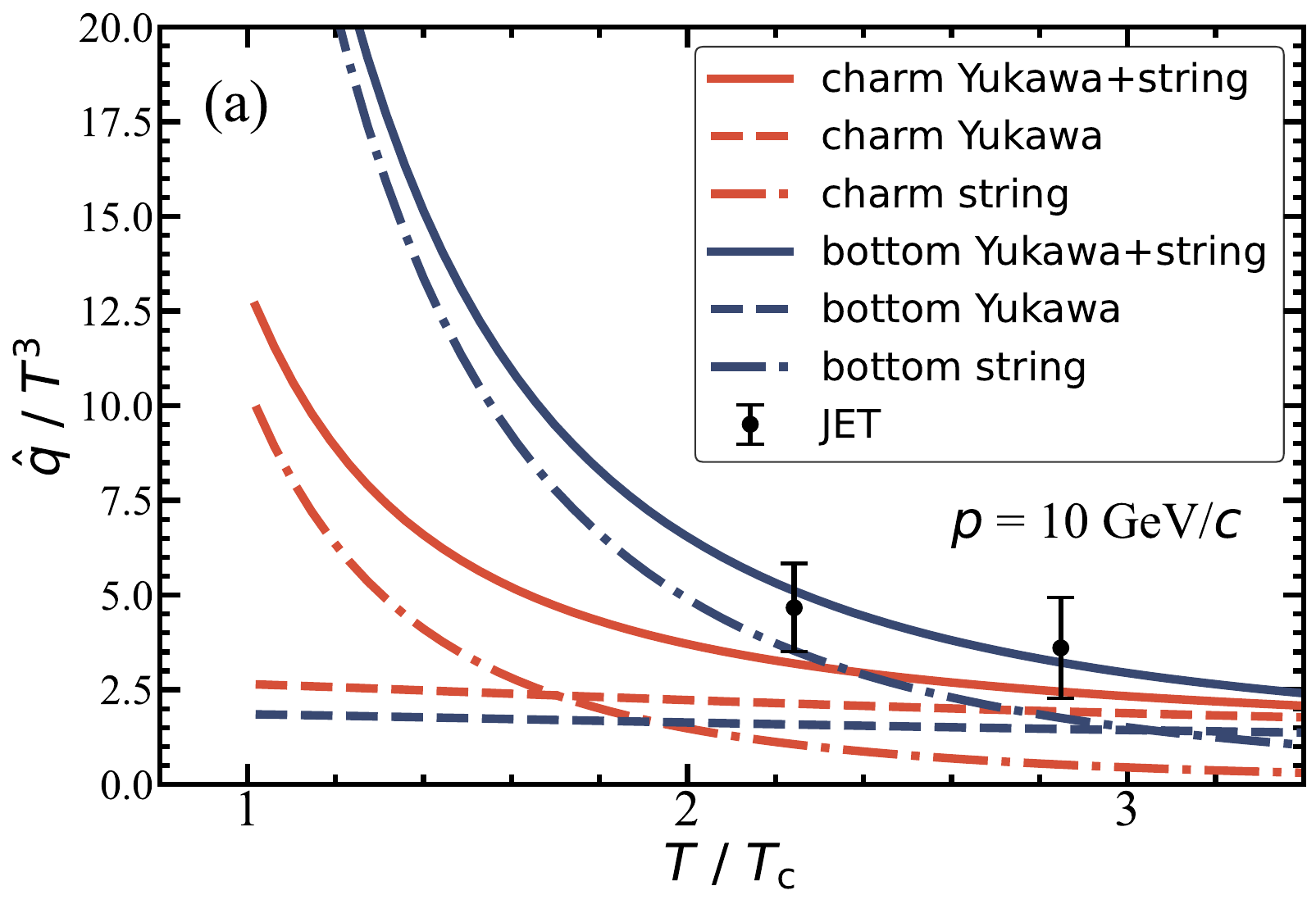}
  \includegraphics[width=0.87\linewidth]{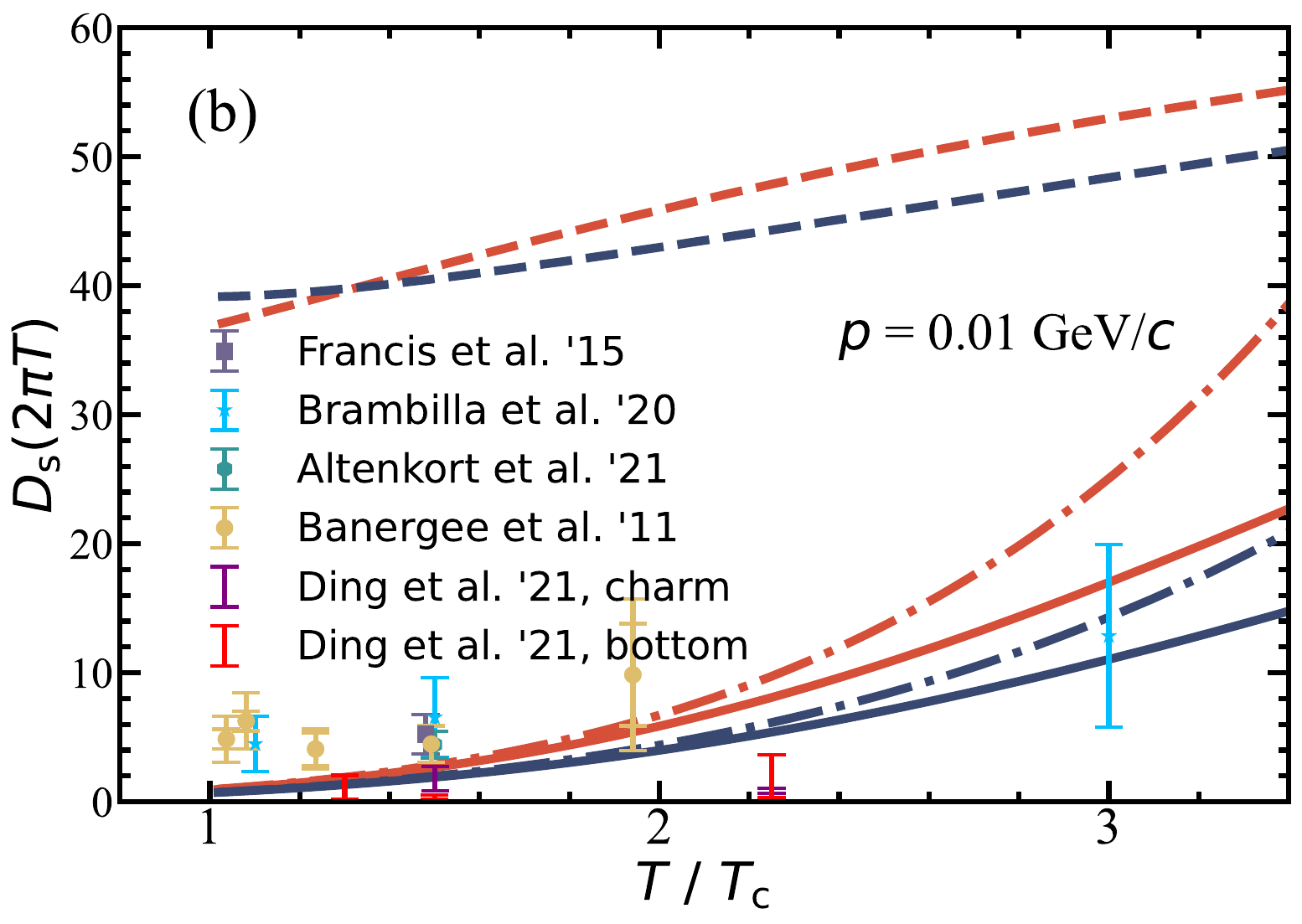}
  \caption{(Color online) The temperature dependences of (a) jet transport coefficient $\hat{q}/T^3$ and (b) spatial diffusion coefficient $D_\mathrm{s}$ of charm and bottom quarks, compared between contributions from Yukawa and string interactions, and also compared to results from the JET Collaboration~\cite{JET:2013cls} and the lattice QCD data~\cite{Banerjee:2011ra,Francis:2015daa,Brambilla:2020siz,Altenkort:2020fgs,Ding:2021ise}.}
  \label{fig:qhatDs}
\end{figure}

To better illustrate Yukawa and string contributions to this non-intuitive hierarchy of heavy quark energy loss at high $p_\mathrm{T}$, we present in Fig.~\ref{fig:qhatp} the $T^3$-rescaled jet transport coefficient as a function of the heavy quark momentum at different temperatures. Comparing between charm and bottom quarks, we notice different hierarchies of their $\hat{q}$ from these two types of interactions. While the Yukawa term gives smaller $b$-quark $\hat{q}$ than $c$-quark $\hat{q}$, as expected from perturbative calculations before~\cite{Cao:2016gvr}, the string term gives larger $b$-quark $\hat{q}$ than $c$-quark $\hat{q}$. This inverse order from the string interaction can be understood with the last term in Eq.~(\ref{eq:M2_cq}) or~(\ref{eq:M2_cg}). Since the Mandelstam variable $t$ is a negative quantity for space-like momentum exchange, this last term increases with the heavy quark mass $m_Q$. This is a qualitative difference from the Yukawa interaction in our model. Comparing between the upper and lower panels of Fig.~\ref{fig:qhatp}, we see while the Yukawa contribution to $\hat{q}$ is not very sensitive to the medium temperature, the string contribution is significantly enhanced when the medium temperature decreases towards $T_\mathrm{c}$. Because the strength of string interaction decays as the momentum exchange between heavy quarks and the QGP becomes larger, and on average, this momentum exchange increases slowly with the heavy quark momentum, the string contribution to $\hat{q}$ becomes weak, though still non-vanishing, for high momentum heavy quarks. Within the temperature and momentum ranges we explore in this work, the total $\hat{q}$ is larger for $b$-quarks than for $c$-quarks, leading to stronger elastic energy loss of $b$-quarks than $c$-quarks. 

Compared to the elastic energy loss, the inelastic energy loss is more complicated. Besides $\hat{q}$, additional mass dependence exists in Eq.~(\ref{eq:gluon_spectrum}), known as the dead cone factor that suppresses the radiative energy loss of heavy quarks. Even though $b$-quark has a larger $\hat{q}$ compared to $c$-quark, the much heavier mass of $b$-quark can still lead to a weaker radiative energy loss when its momentum is not significantly larger than its mass. In the end, the hierarchy of energy loss between charm and bottom quarks depends on the competition between the mass effect on $\hat{q}$ and the dead cone factor, which further rely on the heavy quark momentum and the medium temperature. When the medium temperature is sufficiently high, the string contribution to $\hat{q}$ becomes weak enough to be overcome by the dead cone effect, resulting in larger energy loss of charm quarks than bottom quarks. On the other hand, when the medium temperature is low, the dead cone effect can only defeat the string contribution at low momentum; while at high momentum, the inverse order of energy loss appears. This is exactly what we observe in Fig.~\ref{fig:energyloss}. The hierarchy of $R_\mathrm{AA}$ in realistic heavy-ion collisions depends on the evolution profile of the QGP. For the QGP created in current nuclear collision programs, it stays longer at lower temperature near $T_\mathrm{c}$ than at higher temperature, leading to the crossing of $R_\mathrm{AA}$ between charm and bottom mesons (leptons) seen in Figs.~\ref{fig:RAALO} and~\ref{fig:RAA-LHC}. 

In the end, we present in Fig.~\ref{fig:qhatDs} the transport coefficients of heavy quarks extracted from our model calculation. In the  upper panel, we show the momentum space transport coefficient $\hat{q}/T^3$ as a function of the medium temperature. Consistent with our previous observations in Fig.~\ref{fig:qhatp}, the Yukawa interaction generates larger $c$-quark $\hat{q}$ than $b$-quark $\hat{q}$, while the string interaction generates the opposite order. The Yukawa interaction dominates at high temperature while the string interaction dominates at low temperature. Within the temperature range we explore here, the total value of $\hat{q}$ is larger for $b$-quarks than for $c$-quarks. The values of the heavy quark $\hat{q}$ we obtain here appear consistent with the constraints from the previous JET Collaboration work~\cite{JET:2013cls} on a 10~GeV/$c$ light quark. 

Using the fluctuation-dissipation relation, we can further convert the $\hat{q}$ parameter into the spatial diffusion coefficient of heavy quarks as $D_\mathrm{s}(2\pi T)=8\pi/(\hat{q}/T^3)$~\cite{Cao:2015hia}. In the lower panel of Fig.~\ref{fig:qhatDs}, we present results for 0.01~GeV/$c$ heavy quarks and observe a smaller $D_\mathrm{s}$ for $b$-quarks than for $c$-quarks when contributions from both Yukawa and string interactions are included. This hierarchy was also found in Refs.~\cite{Das:2016llg,Sambataro:2023tlv} within a quasi-particle model, and also seems qualitatively consistent with the hint from the lattice QCD study~\cite{Ding:2021ise} (labeled as ``Ding" in the plot) that evaluates the diffusion coefficients of heavy quarks with finite masses. The values of $D_\mathrm{s}$ we extract for charm and bottom quarks here agree with the range predicted by various lattice calculations~\cite{Banerjee:2011ra,Francis:2015daa,Brambilla:2020siz,Altenkort:2020fgs,Ding:2021ise}. Note that by convention, $D_\mathrm{s}$ is defined for zero momentum heavy quarks. At zero momentum, perturbative calculation using a fixed value of $\alpha_\mathrm{s}$ can also give a smaller $D_\mathrm{s}$ for $b$-quarks than for $c$-quarks~\cite{Liu:2016ysz}, indicating it is harder for heavier particles to diffuse inside a thermal medium. The concept of energy loss is not suitable for these slowly moving heavy quarks. On the other hand, if one extracts $D_\mathrm{s}$ for energetic heavy quarks, e.g., at $p=10$~GeV/$c$, perturbative calculation suggests the value of $D_\mathrm{s}$ is larger for $b$-quarks than for $c$-quarks, indicating weaker energy loss of the former. The momentum at which $b$-quarks and $c$-quarks share the same $D_\mathrm{s}$ depends on the medium temperature and different model assumptions in perturbative calculation, e.g., the values of $\mu_D$ and $\alpha_\mathrm{s}$, and implementation of quantum statistics. In our current LBT model, the possible stronger energy loss of heavier quarks arises from the string interaction.

\section{Summary}
\label{sec:summary}

We have reexamined the mass hierarchy of heavy quark energy loss and their transport coefficients using a linear Boltzmann transport model that incorporates both Yukawa and string interactions between heavy quarks and thermal partons inside the QGP. The general intuition that heavier partons lose less energy inside the QGP has been challenged. We have found that whether bottom quarks lose less or more energy compared to charm quarks depends on the non-trivial interplay between the effect of string interaction on elastic scatterings and the dead cone effect on inelastic scatterings. At low momentum, the strong dead cone effect significantly suppresses the radiative energy loss of massive particles and results in smaller energy loss of bottom quarks than charm quarks as expected. Our model shows larger $R_\mathrm{AA}$ and smaller $v_2$ of $B$ mesons ($b$-decayed leptons) than $D$ mesons ($c$-decayed leptons) as observed in the current RHIC and LHC experiments. To the contrary, at higher momentum, the string interaction that enhances the scattering rate of massive particles can overcome the dead cone effect and generates larger energy loss of bottom quarks than charm quarks. Since the QGP created in realistic heavy-ion collisions spends a considerable portion of its lifetime near the phase transition temperature where the string interaction is strong, one may find smaller $R_\mathrm{AA}$ of bottom particles than charm particles at high $p_\mathrm{T}$. The momentum space transport coefficient ($\hat{q}$) and the spatial diffusion coefficient ($D_\mathrm{s}$) we extract for heavy quarks agree with other phenomenological studies and the lattice QCD data. The string interaction can generate larger $\hat{q}$ and smaller $D_\mathrm{s}$ for bottom quarks than for charm quarks. Although our results here are model dependent, it indicates heavier particles do not necessarily always lose less energy inside the QGP. The upcoming sPHENIX data and more precise measurements at the LHC may provide a more stringent constraint on the string interactions between heavy quarks and a color-deconfined medium.

\section*{Acknowledgments}
We are grateful to valuable discussions with Shuai Y.F. Liu, Feng-Lei Liu, Weiyao Ke, and Shuang Li. This work was supported by the National Natural Science Foundation of China (NSFC) under Grant Nos.~12175122, 2021-867, 12225503, 11890710, 11890711, and 11935007.

\bibliographystyle{h-physrev5}
\bibliography{SCrefs}

\end{document}